# On-chip non-reciprocal optical devices based on quantum inspired photonic lattices

R. El-Ganainy[1,2,*], A. Eisfeld[1], Miguel Levy[2] and D.N. Christodoulides[3]

[1]Max Planck Institute for the Physics of Complex Systems, Nothnitzer street 38, 01187 Dresden, Germany

[2]Department of Physics, Michigan Technological University, Houghton, Michigan 49931

[3]College of Optics & Photonics-CREOL, University of Central Florida, Orlando, Florida

*Corresponding author: ganainy@mtu.edu

We propose a novel geometry for integrated photonic devices that can be used as isolators and polarization splitters based on engineered photonic lattices. Starting from optical waveguide arrays that mimic Fock space representation of a non-interacting two-site Bose Hubbard Hamiltonian, we show that introducing magneto-optic nonreciprocity to these structures leads to a superior optical isolation performance. In the forward propagation direction, an input TM polarized beam experiences a perfect state transfer between the input and output waveguide channels while surface Bloch oscillations block the backward transmission between the same ports. Our analysis indicates a large isolation ratio of 75 dB after a propagation distance of 8 *mm* inside seven coupled waveguides. Moreover, we demonstrate that, a judicious choice of the nonreciprocity in this same geometry can lead to perfect polarization splitting.

Integrated optics is playing an ever increasingly important role in modern technology. From communication networks to medical applications, photonic devices are becoming vital elements in every commercial system where their different functionalities are enabled by variety of physical and engineering concepts. The viability of these photonic devices depends on several important parameters such as fabrication tolerances, compatibility with existing technologies, power management and versatility. While the last decade has witnessed rapid developments in miniaturizing photonic components, realization of high performance commercial on-chip optical isolators still remains a hurdle. These devices operate as optical diodes that allow unidirectional transmission of light and they are essential components in most optical systems.

Proper operation of optical isolators necessitates large isolation ratios of at least 50 dB, low insertion loss and negligible optical absorption. In order to satisfy the aforementioned requirements, today's commercial isolators are based on Faraday rotation [1]. However, these devices suffer from the major drawback of being bulky and incompatible with integrated optical platforms. Attempts to overcome this difficulty have led to a surge of new ideas and concepts for designing integrated optics isolators. These investigations range from using integrated magnetooptics material [2,3], acoustic/electro-optics effects [4] and Kerr nonlinearity [5] to non-Hermitian optics [6]. Optical diodes based on nonreciprocal silicon rings have been also reported [7].

Recently a new optical isolator device based on unidirectional optical Bloch oscillations (BO) was proposed and 35 dB isolation ratios were shown in simulations [8-10]. Following this work, it was shown that non-reciprocal resonant delocalization (RD) effects [11] can lead to an even higher performance (45 dB) and smaller footprint [12]. Despite the characteristic advantages and limitations of each of the above techniques, the non-reciprocal Bloch oscillations are emerging as a very promising paradigm for realizing on-chip optical diodes. For instance it is compatible with mature silicon-on-insulator (SOI) technologies [8-10] and does not depend on any nonlinear processes that might degrade the performance at low input powers. In addition, it is based on waveguide transmission structures and thus it enjoys a large bandwidth of operation [13]. Moreover, careful waveguide coupling can minimize insertion loss in these geometries. Finally we note that their operation requires small magnetic effects (compared with those used in commercial isolators based on Faraday rotation based) that can be provided using thin film magnetic garnets [8].

In this letter, we propose high performance integrated optical isolators based on quantum inspired uniform non-reciprocal waveguide lattices with inhomogeneous coupling constants. In contrast to [12], the proposed geometry does not involve modulation of the waveguide width along the propagation direction and is thus easier to fabricate. In addition it achieves a much superior performance to those reported in [8,12]. Even more importantly, we demonstrate the versatility of our proposed device by showing that a slight modification of the design leads to a totally different functionality: that of a polarization splitter. The operation of this photonic diode is based on a complete transfer of optical intensity from one waveguide site at the input to another channel at the output during forward propagation [14]. On the other hand, surface revival effects block any backward propagation to the input waveguide element, thus providing optical isolation [14].

In order to illustrate the enabling concepts behind the proposed optical isolator, we start by describing how to synthesize the waveguide arrays employed in these devices. We do so by considering a 2-site non-interacting Bose-Hubbard Hamiltonian:

$$H = \hbar \beta_a \hat{a}^+ \hat{a} + \hbar \beta_b \hat{b}^+ \hat{b} + \hbar \kappa \left( \hat{a}^+ \hat{b} + \hat{a} \hat{b}^+ \right) \quad (1)$$

This Hamiltonian can model many diverse physical systems. For instance it describes the propagation of non-classical light in an unbalanced directional coupler [15] as well as the evolution of quantum light in detuned optical cavities as depicted in Figs.1(a) and (b) respectively. Despite its simplicity, the Hamiltonian in Eq.(1) was shown to exhibit a host of intriguing behavior ranging from surface Bloch oscillations to discrete supersymmetry (SUSY) [14].

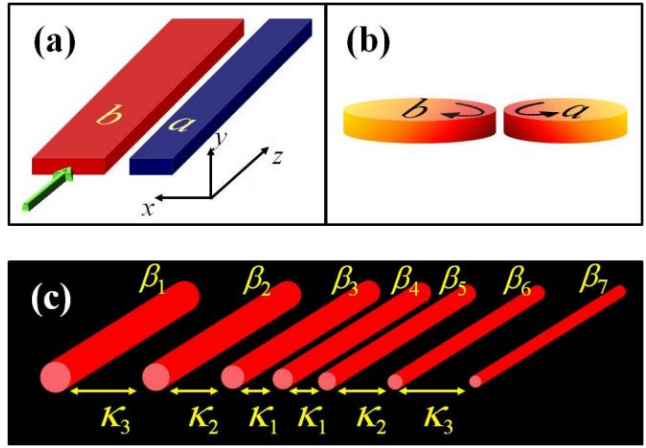

Fig. 1. A schematic of nonclassical light in (a) two coupled waveguides and (b) two coupled cavities. (c) depicts the optical analogue of Fock space representation of 6 photons propagating in the structure shown in (a).

Here we focus our analysis on the directional coupler scenario. Accordingly, $\kappa$ is the coupling coefficient between waveguides having propagation constants of $\beta_{a,b}$. In Eq.(1), $\hat{a}^+$ and $\hat{a}$ are the bosonic creation and annihilation operators of photons in waveguide $a$ while $\hat{b}^+$ and $\hat{b}$ are those associated with waveguide $b$. We stress that the Hamiltonian in eq.(1) does not describe the isolator structure but is rather used as a mathematical machinery to synthesize the waveguide array that will be used later as an isolator  The Hamiltonian $H$ can be directly diagonalized by means standard unitary transformation [15] in which case we obtain:

$$H = \hbar\beta_e \hat{c}_e^+ \hat{c}_e + \hbar\beta_o \hat{c}_o^+ \hat{c}_o \quad , \quad (2)$$

where $\hat{c}^+_{e,o}$ and $\hat{c}_{e,o}$ are the bosonic creation and annihilation operators of the even/odd-like eigenstates of equation (1) and $\beta_{e,o} = \frac{\beta_a + \beta_b}{2} \pm \sqrt{\left(\frac{\beta_a - \beta_b}{2}\right)^2 + \kappa^2}$.

We now consider a state having $2N$ light quanta $|\psi_{N,n}\rangle$ with $N \mp n$ occupying even/odd modes, respectively: $|\psi_{2N,n}\rangle = \frac{(\hat{c}^+_e)^{N-n}(\hat{c}^+_o)^{N+n}}{(N-n)!(N+n)!}|vac\rangle = |N-n, N+n\rangle_c$. where $|n| \le N$ and the subscript $c$ is used to indicate that these states are in the even/odd basis. It is straightforward to show that $\{|\psi_{2N,n}\rangle\}$ are eigenstates of the Hamiltonian operator $H$ with corresponding eigenvalues $\lambda_{2N,n} = \hbar\{N(\beta_o + \beta_e) + n(\beta_o - \beta_e)\}$ distributed on a ladder of step equal to $\hbar(\beta_o - \beta_e)$ [14].

Equation (1) can be also equivalently analyzed in more natural bases described by with $N \mp n$ photons in the left/right waveguide channels respectively: $|n_{2N}\rangle = \frac{(\hat{a}^+)^{N-n}(\hat{b}^+)^{N+n}}{(N-n)!(N+n)!}|vac\rangle = |N-n, N+n\rangle$. In this representation, the evolution of an arbitrary wavefunction $|\varphi(z)\rangle = \sum_{n=-N}^{N} \exp(-2iN\beta_{avg}z)\chi_n(z)|n_{2N}\rangle$ with a given set of initial conditions $\chi_n(0)$ is given by:

$$i\frac{d\chi_{-N}}{dz} = -\Delta_{ba}N\chi_{-N} + \kappa\sqrt{2N}\chi_{-N+1}$$
$$i\frac{d\chi_n}{dz} = \Delta_{ba}n\chi_n + \kappa(g_n\chi_{n-1} + g_{n+1}\chi_{n+1}) , \quad -N < n < N \qquad (3)$$
$$i\frac{d\chi_N}{dz} = \Delta_{ba}N\chi_N + \kappa\sqrt{2N}\chi_{N-1}$$

Where $g_n = \sqrt{(N+n)(N-n+1)}$, $\beta_{avg} = (\beta_a + \beta_b)/2$ and $\Delta_{ba} = \beta_b - \beta_a$.

Equation (3) represents the Fock space dynamics of $H$ and in general can be simulated by a classical array of coupled waveguides [14]. Interestingly, Eq.(3) can be also derived from the consideration of $N/2$ spin particles [16]. This is by no means an accident. In fact the two pictures can be reconciled within the frame work of Schwinger boson formalism of spin and angular momentum operators [17].

While perfect state transfer (complete transfer of excitation from one input channel to a different element at the output) was proposed in spin lattices [16] and later demonstrated in analogous photonic systems for $\Delta_{ba} = 0$ [18], surface revivals (the spread and refocusing of an input beam launched at the edge of the array) and Bloch oscillations were recently predicted for finite non-zero detuning ($\Delta_{ba} \neq 0$) [14] and experimentally demonstrated for semi-infinite Glauber-Fock photonic arrays [19]. We note that both effects (state transfer and surface revivals) are a direct outcome of the equidistant distribution of the eigenvalues of $H$ together with some symmetry constraints [14,16].

Here we propose to exploit the above mentioned features to design high performance integrated optical isolators as well as on-chip polarization splitters. In order to do so, we consider a waveguide array equivalent to that described by eq.(3) and shown in Fig.1(c). Here we consider a realistic array design consisting of silicon-on-insulator ridge waveguides similar to those schematically shown in Fig.(2). The descending order of the waveguide width from left to right indicates a linear ramp of the propagation constant. Note that this linear ascendance of the propagation constants does not necessarily correspond to a linear gradient in waveguide widths. Moreover, if necessary, waveguide heights can be used as an additional degree of freedom to achieve the system requirements. Finally breaking time reversal symmetry between forward and backward propagation can be introduced by depositing magnetic garnet (bismuth or cerium-substituted yttrium iron garnets) films on top of each ridge waveguide [8-10]. The thickness of silicon ridge channels determines the strength of the nonreciprocity associated with that particular waveguide element. Alternatively a composition gradient in the magnetic garnet cover layer may also be used to produce a nonreciprocity gradient across the array. The only difference between the waveguide array of Fig.(2) and those studied in [8] is the engineered distance between the elements to provide inhomogeneous coupling coefficients. Figures 2 (a) and (b) demonstrate the operation of such device as an isolator. The magnetic garnet films (not shown here) and the applied magnetic fields are designed such that the propagation constants are uniform in the forward direction while the linear ramp in the effective index persists in the backward propagation. Detailed design parameters used to achieve this functionality are discussed in ref. [8].

For concreteness we consider a system made of 7 waveguide elements with $N=3$ and $\kappa = 7/\sqrt{12}\ cm^{-1}$ operating at the telecommunication wavelength of $\lambda = 1.55\ \mu m$. Consequently, the coupling coefficients of the array take the values: $7 \times \begin{bmatrix} 1/\sqrt{2} & \sqrt{5/6} & 1 & 1 & \sqrt{5/6} & 1/\sqrt{2} \end{bmatrix} cm^{-1}$. Finally, for this particular example, the backward ramp in the propagation is chosen as $\Delta_{ba} = 0.82 \times 7 = 5.74\ cm^{-1}$.

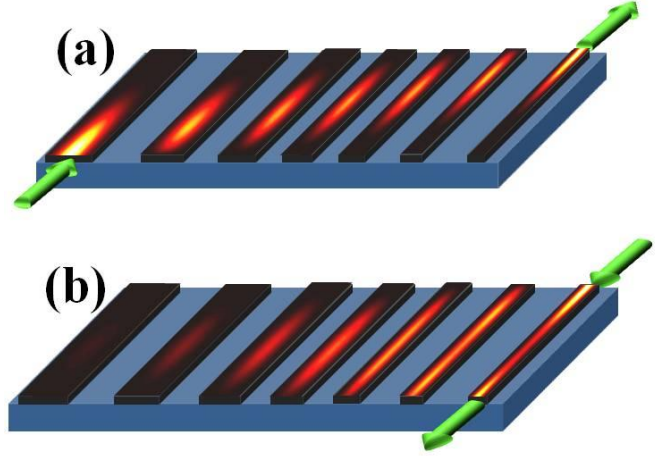

Fig.2. A schematic of the proposed optical isolator structure and its operation. (a) Complete state transfer is achieved in the forward direction and an input optical beam in the left most waveguide will exit from the right most channel after 7.8 *mm* of propagation. (b) Surface revival effects dominate in the backward propagation and high isolation ratio of 75 dB is predicted. The array design parameters are discussed in the text.

As shown in Fig.2(a), an input beam launched at the left most waveguide will undergo a complete state transfer [18] (all optical power is transferred to the right most waveguide) after a propagation distance given by $L_t = \pi/(2\kappa) \approx 7.8\,mm$. On the other hand, a backward reflected beam will experience an effectively 'tilted' array due to the propagation constant ramp. As a result surface revival effects take place with a revival period of $L_r = \pi/\sqrt{\left(\frac{\beta_a - \beta_b}{2}\right)^2 + \kappa^2} \approx 9\,mm$.

Under these conditions, most of the backward optical power exits from the right most waveguides and a large isolation ratio of ~75 dB is achieved.

We note that this is a dramatic improvement over the results obtained before using waveguide arrays with uniform coupling. Moreover, our analysis indicates a profound performance when $\Delta_{ba} = 7\,cm^{-1}$. In the latter scenario, all the backward optical power exits from the right most waveguide with null output from the left most channel and a perfect isolation is achieved. Note however that this requires a 22% increase in the degree of the nonreciprocity. In practice, 75 dB isolation is very large and together with a smaller degree of nonreciprocity might be preferable to a perfect isolation.

The prediction of such a high degree of isolation elevates the upper limit on the isolator performance. It also indicates that given a suitable mechanism for controlling the garnet films magnetism, one might be able to use the device as a switch to direct input beams to either one of the two output channels (left/right most waveguides). This merits further investigation and we will carry out this study elsewhere.

Finally we now illustrate that this perfect isolation regime can be used for polarization splitting. In order to do so we note that while all the previous analysis applies only for TM polarized light [8-10], TE modes are not affected by the magnetization.

Let us now consider an input beam composed of a superposition of TM/TE modes. This scenario is depicted in Fig.3. Note that, for illustration purposes, the propagation of the two superimposed polarizations is separated into two sub-plots. As shown in the bottom (green) part of Fig.3, the TM mode 'sees' a uniform lattice and its optical power is fully transferred to the opposite side of the array. On the other hand, the TE mode will be subject to the lattice 'tilt' and surface revivals takes over. Thus the output TE power exits from the same input channel as depicted in the upper part of Fig.3. As a result, a perfect splitting between the two states of polarization is achieved.

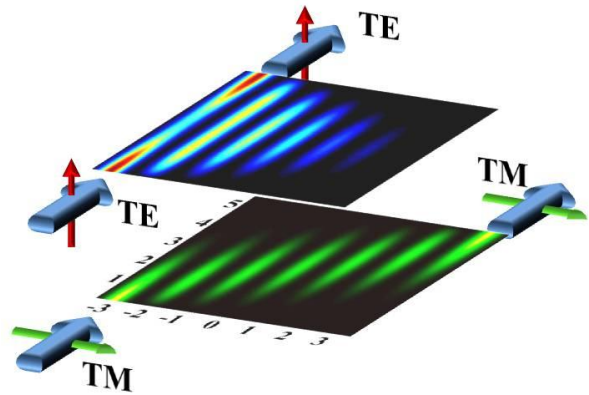

Fig.3. On-chip polarization splitter. An optical beam composed of two orthogonal TE/TM polarizations launched in the left most waveguide element. Due to its sensitivity to the magneto-optic effects, The TM mode experiences a uniform array and undergoes a perfect state transfer shown in the bottom part of the figure. On the other hand, the TE polarization can sense the lattice tilt and exits from the same input channel as shown on the top figure. Small arrows (in red and green) indicate the transverse component of the magnetic field of the modes.

We note that several proposals for TE/TM splitting have been suggested and demonstrated in the literature ( see for example [20,21] ). The obvious advantages of the design presented here lie in its high performance and its compatibility with optical isolator designs. In other words, the same structure can have dual functionality. This last feature is extremely important and provides a tremendous advantage for integrated optics industry, namely that same fabrication techniques and facilities can be used without any modifications to build two different devices. Hence, not only our proposed design provides an avenue for high performance devices but it also offers a great opportunity for low cost fabrication.